\documentstyle[psfig,conf_iap,]{article}
\newdimen\ww
\begin{document}
\heading{
The gaseous content 
of the Blue Compact 
Galaxy Mrk~59} 
\par\medskip\noindent
\author{
Alain Lecavelier des Etangs$^{1}$, Trinh X. Thuan$^{2}$, Yuri I. Izotov$^{3}$
}
\address{
Institut d'Astrophysique de Paris, CNRS, 98bis Boulevard Arago, 
       F--75014 Paris}
\address{
Astronomy Department, University of Virginia, Charlottesville, VA 22903
}
\address{
Main Astronomical Observatory, National Academy of Sciences, 03680 Kyiv, 
Ukraine}

\begin{abstract}
{\sl FUSE} far-UV spectroscopy of the nearby metal-deficient ($Z_\odot$/8) 
cometary Blue Compact Dwarf (BCD) galaxy Mrk59 have been 
used to investigate element abundances in its interstellar medium.
The lack of diffuse H$_2$ ($N$(H$_2$)$\le 10^{15}$cm$^{-2}$) is due to 
the combined effect of a strong UV radiation field and a low metallicity.
By fitting the line profiles with multiple components having 
$b$ = 7 km s$^{-1}$ and spanning a radial velocity 
range of 100 km s$^{-1}$, we derive 
heavy element abundances in the neutral gas. 
We find log $N$(O~{\sc i})/$N$(H~{\sc i})
 = -- 5.0$\pm$0.3, 
about a factor of 10 below the oxygen abundance of the supergiant 
H~{\sc ii} region.

\end{abstract}

\section{FUSE Observations of Mkr~59}

The Blue Compact Dwarf (BCD) galaxy Markarian 59 (Mrk 59) $\equiv$ I Zw 49 
belongs to the 
class of cometary BCDs with an intense starburst at the end of an elongated 
low surface brightness stellar body. 
This BCD have numerous knots composed of a chain of 
H~{\sc ii} regions, resulting from propagating star formation along 
the galaxy's elongated body and ending with the high-surface-brightness 
supergiant H~{\sc ii} region.
From deep ground-based spectrophotometric observations  
of the supergiant H~{\sc ii} regions Noeske et al. (2000) 
derived an Oxygen abundance log O/H = -4.011 $\pm$ 0.003 
($Z_\odot$/8).
O abundances were also derived for two other emission knots along the elongated 
body and were found to be the same within the errors. The small scatter in metallicity 
along the major axis of Mrk 59 ($\sim$ 0.2 dex)
suggests that the mixing of elements in the ionized gas 
has been efficient on a spatial 
scale of several kpc. 

Mrk 59 was observed during 7,865 s on 2000, January 11 with {\sl FUSE} 
(Moos et al. 2000). The LWRS large entrance 
aperture (30"$\times$30") was used, so that all of Mrk 59 
which is $\sim$ 20" across (Noeske et al. 2000) is included within it.

Mrk 59 shows numerous interstellar absorption lines from the H~{\sc i} Lyman series 
(Ly $\beta$ to Ly $\lambda$ (H~{\sc i} 11))
and from other atoms and ions such as C~{\sc ii}, C~{\sc iii}, N~{\sc i}, 
N~{\sc ii}, N~{\sc iii}, O~{\sc i}, Si~{\sc ii}, S~{\sc iii}, 
S~{\sc iv}, Fe~{\sc ii} and Fe~{\sc iii}. 
Several stellar features such as the Si~{\sc iv} and S {\sc vi} lines 
are also detected (Thuan et al. 2001a).

\section{Narrow cores and broad wings of the H~{\sc i} Lyman series   
line profiles: interstellar and  
stellar absorption}

We found that the shapes of the H~{\sc i} Lyman series lines 
cannot be reproduced with Voigt profiles 
characterized by a single H~{\sc i} column density. 
Damping wings are clearly 
visible, signaling a very large H~{\sc i} column density 
$\sim 5\times 10^{22}$~cm$^{-2}$.   
However, this high column density is not consistent with the shape of the
strongest H~{\sc i} lines (Lyman~$\beta$ to Lyman~$\delta$). 
For instance, although damping wings are also present, the core of the Lyman~$\delta$ line 
is thin and cannot be fitted with a model profile with  
the column density needed to fit the higher lines in the series.
The thin core implies a smaller column density, of the order of 
$10^{21}$~cm$^{-2}$.
We interpret that by a non-interstellar but stellar origin of the wide damping wings.
We conclude that 
the broad wings of the above lines 
arise in the photospheres of numerous B stars,
while their narrower cores are caused by interstellar absorption.
By fitting the cores of the Lyman series lines, we obtain an 
interstellar H~{\sc i} column density of 
$\sim 7 \times$10$^{20}$~cm$^{-2}$.

\section{Upper limits on the diffuse H$_2$ content of Mrk 59}

No line of H$_2$ is seen at the radial velocity of Mrk~59.
Using nine H$_2$ Lyman bands (0--0 to 8--0),
we conclude that the total column density of diffuse H$_2$ is 
$\le$10$^{15}$cm$^{-2}$.
This implies that the ratio of H$_2$ to H{\sc i} is $\le$10$^{-6}$. 
The corresponding average molecular fraction
$f$=2$N$(H$_2$)/($N$(H~{\sc i})+2$N$(H$_2$)) is $\le 3 \times 10^{-6}$.

By calculating the amount of H$_2$ molecules formed on  
the surface of dust grains, we find that 
the expected $N$(H$_2$) must be $\sim$2$\times$10$^{12}$~cm$^{-2}$, 
consistent with our 
upper limit. As compared to the Milky Way, the low column density 
of diffuse H$_2$ in Mrk 59 
is due to the combined effects of a large UV flux which destroys the 
H$_2$ molecules and of a low metallicity resulting in 
a scarcity of dust grains on which H$_2$ form (see the full discussion
in Thuan et al. 2001a).   

\section{Heavy element abundances}    

To estimate the abundances, we consider two cases. 
The first case presented during the meeting assumes that there is a single 
velocity component along the thousands of lines of sight to the stars of Mkr~59. 
In this case, absorption lines 
that are broader than the instrumental Point Spread Function
and do not go down to 
zero are supposed to be not saturated.
The second case, suggested during the meeting (York, public communication),
assume multiple velocity components along the multiple lines of sight, 
some of which may have saturated profiles. 

\begin{center}
\begin{table}
\caption{Heavy element column densities in Mrk 59}
\begin{tabular}{lrcc|lrcc}
\hline
 &$b_1$$^{\rm a}$
& log $N_1$$^{\rm a}$
& log $N_2$$^{\rm b}$ & 
 &$b_1$$^{\rm a}$
& log $N_1$$^{\rm a}$
& log $N_2$$^{\rm b}$\\
\hline
C~{\sc iii}  &$\leq$ 35 & 18.4$\pm$0.1  & 18.2$\pm$0.1 &
N~{\sc i}    &40$\pm$20 & 14.0$^{+0.3}_{-0.4}$  & 14.6$^{+0.4}_{-0.3}$\\
N~{\sc ii}   &30$\pm$10 & 14.2$^{+0.1}_{-0.2}$ & 15.1$^{+2.2}_{-0.8}$ &
O~{\sc i}    &40$\pm$20 & 15.1$^{+0.3}_{-0.3}$ & 15.8$^{+0.3}_{-0.3}$ \\
S~{\sc iii}  &95$\pm$15 & 15.1$\pm$0.1 & 19.0$^{+0.1}_{-0.2}$         &
S~{\sc iv}   &30$\pm$10 & 14.2$^{+0.1}_{-0.2}$ & 14.9$^{+0.9}_{-0.6}$ \\
Fe~{\sc ii}  &40$\pm$15 & 13.9$^{+0.1}_{-0.2}$ & 14.4$^{+0.3}_{-0.3}$&
Fe~{\sc iii} &105$\pm$10& 15.1$\pm$0.1 & 18.9$^{+0.1}_{-0.1}$\\
\hline
\end{tabular}
a){Fitting with a single Voigt profile. $b_1$ is in km\,s$^{-1}$.
Error bars are 3 $\sigma$ limits.}\\
b){Fitting with multiple velocity components, 
each with $b_2$ = 7 km s$^{-1}$, and spanning a velocity 
range of 100 km\,s$^{-1}$.  Error bars are 2 $\sigma$ limits.}\\
\vspace{-0.5cm}
\end{table}
\end{center}

\vspace{-1.cm}

\subsection{Profile fitting with a single velocity component}      

We consider first the single interstellar velocity component case.
Some of the lines, like the
Fe\,{\sc iii} and S\,{\sc iii} lines, are very broad
with unreasonably large 
Doppler widths ($b$$\ge$100~km~s$^{-1}$), 
suggesting, as for the H{\,\sc i} lines, that 
these metal lines have some contamination by stellar absorption.
The resulting column densities ($\log N_1$) and Doppler widths ($b_1$) 
along with their error bars  
are given in Table~1. The error bars of the column densities 
include the uncertainty in $b$. 
These abundances of the metals relative to 
$N$(H\,{\sc i}) are extremely low.
In particular, with log $N$(O\,{\sc i})/$N$(H\,{\sc i}) = -- 5.7,
the oxygen abundance in the H\,{\sc i} gas could be $\sim$ 50 times lower  
than the oxygen abundance in the ionized gas, as determined from the optical 
emission line spectrum of the supergiant H\,{\sc ii} region in Mrk 59. 
However, these very low abundances may be underestimates due to 
the saturation of some lines of sight.

\subsection{Profile fitting with multiple velocity components}

We are observing through the {\sl FUSE} aperture
along thousands of lines of sight towards UV-bright stars.   
The observed spectrum is the sum of many spectra.
It can happen that some lines of sight have saturated absorption lines
with a small $b$ parameter,
but because they have different radial velocities spread 
over several tens of km\,s$^{-1}$, 
the broad absorption line resulting from the sum of many narrow absorption 
lines does not go to zero intensity, and its  
width is larger than the Point Spread Function. In that case, a 
single velocity component fit to the line profile  
would result in an overestimate of 
the $b$ parameter and in an underestimate of the column density.

\begin{figure}[tbh]
\centerline{\vbox{
\psfig{figure=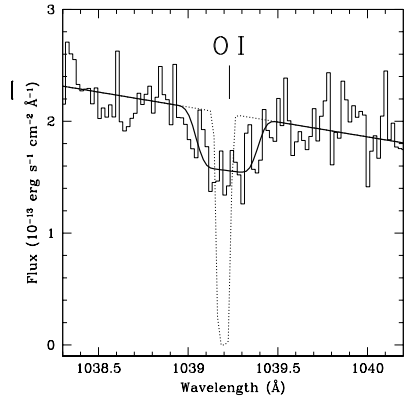,height=5.5cm}
}}
\caption[]{Fit to the O~{\sc i} 1039\AA\ absorption line with a
column density $N$(O~{\sc i}) = 7 $\times$ 10$^{15}$cm$^{-2}$. 
The fit is obtained by the addition of multiple profiles.
$\Delta$$v$ and $b$ parameters are constrained by the simultaneous fit of 
several lines of O~{\sc i}, N~{\sc i} and Fe~{\sc ii}.
The thick solid line gives the resulting fit to the data with 
$\Delta$$v$ = 100 km~s$^{-1}$ and $b$ = 7 km s$^{-1}$.
The dotted line represents an example of one line of sight with 
$b$ = 7 km s$^{-1}$. With this column density, such an 
individual line of sight is
saturated and goes down to the zero intensity level. 
However the absorption line resulting from the 
addition of many narrower profiles does not go to the 
zero intensity level, although it is broader than the width of the 
instrumental Point Spread Function ($\sim$ 0.1\AA).
\vspace{-.5cm}}
\end{figure}

To investigate this issue, we have calculated profiles resulting from the
addition of multiple lines of sight. We adopt the simple model where  
the different lines of sight have radial velocities distributed 
uniformly between $v_{\rm Mrk 59} - {\Delta}v/2$ and $v_{\rm Mrk 59} + 
{\Delta}v/2$, where ${\Delta}v$ is the spread in velocity due mostly 
to the velocity dispersion of the system of 
absorbing clouds along the multiple lines of sight, and also partly to the
wavelength smearing caused by the extension of Mrk 59 within the aperture.
We determine the best estimates of $b$ (assumed to be the same for
all lines of sight) and ${\Delta}v$ by the simultaneous fit of several lines
of O~{\sc i}, N~{\sc i} and Fe~{\sc ii}.
Each of these elements having several transitions with different oscillator
strengths, this well constrains $b$ and $\Delta v$ to reasonable values: 
$b$=7$^{+13}_{-3}$\,km\,s$^{-1}$ and 
${\Delta}v$=100$\pm$20\,km\,s$^{-1}$ (2$\sigma$ limits, see Fig~1).
Note that the value of ${\Delta}v$ is 
very close to the FWHM of 92\,km\,s$^{-1}$ of 
the H~{\sc i} profile (Thuan et al. 2001b). 
The heavy element column densities derived with the multi-component 
fit are given in Table\,1 ($\log N_2$).
Those are larger than in the case of a single 
component fit, by a mean factor of $\sim$5. The very large column densities 
derived for C\,{\sc iii}, S\,{\sc iii} and Fe\,{\sc iii} are caused by 
stellar contamination. It is interesting to note that,
even with a higher derived oxygen abundance, log $N$(O~{\sc i})/$N$(H~{\sc i})
 = -- 5.0$\pm$0.3, 
the H~{\sc i} absorbing cloud 
still has a metallicity lower than that of the supergiant H~{\sc ii} region 
by a factor of $\sim$ 10. This suggests self-contamination of 
the H~{\sc ii} region by heavy elements released during the present burst of 
star formation (Kunth \& Sargent 1986). 
While mixing of these newly formed heavy elements 
appears to have occurred on the scale ($\sim 2$~kpc) of the H~{\sc ii} regions
(Noeske et al. 2000), 
it has not had time to occur for the whole neutral gas component as the H~{\sc i}
envelope surrounding the star-forming regions is much more extended 
(its diameter is $\sim$ 19 kpc from the VLA map by Thuan et al. 2001b). 

\acknowledgements{
The profile fitting was done using the Owens procedure  
developed by Dr.~M.~Lemoine and the {\sl FUSE} French Team.
We warmly thank Dr.~D.~York for his important suggestion during the meeting.}

\vfill
\end{document}